\documentclass[runningheads]{llncs}
\usepackage[T1]{fontenc}
\usepackage{graphicx,verbatim}
\usepackage{amssymb}
\usepackage{marvosym}
\begin{document}
\title{From Trajectories to Phenotypes:\\
Disease Progression as Structural Priors for Multi-organ Imaging Representation Learning}
\titlerunning{From Trajectories to Phenotypes}
%

\author{Zian Wang\inst{1} \and
Lizhen Lan\inst{2,7} \and
Guangming Wang\inst{2} \and
Haosen Zhang\inst{2} \and
Minxuan Xu\inst{3} \and
Qing Li\inst{2} \and
Tianxing He\inst{2} \and
Mo Yang\inst{2} \and
Wenyue Mao\inst{4} \and
Yajing Zhang\inst{5} \and
Yan Li\inst{6}\textsuperscript{({\Letter})} \and
Chengyan Wang\inst{2}\textsuperscript{({\Letter})}}
\authorrunning{Zian Wang et al.}
\institute{College of Computer Science and Artificial Intelligence, Fudan University, Shanghai, China \and
Human Phenome Institute, Fudan University, Shanghai, China \\
\email{wangcy@fudan.edu.cn} \and
College of Intelligent Robotics and Advanced Manufacturing, Fudan University, Shanghai, China \and
Institute of Science and Technology for Brain-Inspired Intelligence, Fudan University, Shanghai, China \and
Science and Technology Organization, GE Healthcare, Beijing, China \and
Department of Radiology, Ruijin Hospital, Shanghai Jiao Tong University School of Medicine, Shanghai, China \\
\email{ly40730@rjh.com.cn} \and
Digital Medical Research Center, School of Basic Medical Sciences, Fudan University, Shanghai, China}
  
\maketitle              
\begin{abstract}
Imaging-derived phenotypes (IDPs) summarize multi-organ physiology but provide only static snapshots of diseases that evolve over time. In contrast, longitudinal electronic health records encode disease trajectories through temporal dependencies among past diagnosis events and comorbidity structure. We hypothesize that IDPs and disease trajectories contain partially shared disease-relevant structure.
We propose a trajectory-aware distillation framework that transfers structural knowledge from a generative disease trajectory Transformer into an organ-wise IDP encoder. A population-scale trajectory model trained on longitudinal diagnosis sequences produces subject-level embeddings that supervise IDP representation learning via geometry-preserving alignment. During downstream prediction, trajectory and imaging representations can also be fused via cross-attention.
Across 159 diseases in the UK Biobank cohort, trajectory-aware pretraining consistently improves both discrimination (AUC) and time-to-onset prediction (MAE), with the largest gains for low-prevalence diseases. Similarity relationships in IDP embedding space also align with those in trajectory space, providing supportive evidence for partially aligned representation geometry.
These results suggest that population-scale generative disease models can serve as structural priors for data-limited imaging modalities, improving robustness under realistic cohort constraints.
\keywords{Disease Trajectory \and Knowledge Distillation \and Multi-organ \and Imaging-derived Phenotypes}

\end{abstract}
%
%
%

%
\section{Introduction}

Imaging-derived phenotypes (IDPs) provide quantitative summaries of multi-organ medical images and are promising for disease risk modeling in large cohorts such as the UK Biobank (UKB)~\cite{bycroft2018,miller2016}. However, they remain static snapshots of an inherently dynamic disease process. Because disease progression depends on temporal diagnosis patterns and comorbidity structure, IDP-only representations may miss aspects of progression geometry, especially when imaging cohorts are limited.

In contrast, longitudinal electronic health records (EHRs) can capture disease evolution at population scale. Recent transformer-based models, including BEHRT~\cite{li2020behrt}, Med-BERT~\cite{rasmy2021medbert}, and generative disease trajectory models~\cite{shmatko2025}, show that diagnosis sequences can yield informative trajectory embeddings. We hypothesize that these embeddings provide structural priors that can guide imaging representation learning~\cite{bengio2013representation,hinton2015distillation}.

Motivated by this hypothesis, we propose a \textbf{trajectory-aware distillation framework} that transfers disease progression knowledge from a pretrained generative trajectory model into an imaging-based disease prediction model. A trajectory transformer trained on large-scale diagnosis records provides subject-level embeddings, which are used to pretrain an organ-wise IDP encoder via representation alignment~\cite{radford2021clip,liu2025multimodal,tian2020crd}. This encourages IDP embeddings to preserve disease-relevant structural information learned from longitudinal data. We further integrate trajectory and imaging representations using cross-attention fusion for downstream disease prediction. Beyond improving predictive performance, we show that similarity relationships in IDP embeddings reflect similarity in disease trajectories, which we interpret as supportive evidence for partially shared disease structure.

\textbf{Our contributions are summarized as follows:}

\textbf{(1) Trajectory-aware distillation for imaging representation learning.}
We introduce a distillation framework that transfers population-scale disease trajectory knowledge into IDP embeddings.

\textbf{(2) Improved disease prediction through trajectory-informed pretraining and fusion.}
We demonstrate consistent improvements across 159 diseases, especially for low-prevalence conditions.

\textbf{(3) Supportive evidence for shared structure between trajectory and imaging representations.}
We show that similarity relationships in IDP embeddings reflect disease trajectory structure, providing supportive evidence for partially shared disease-relevant structure.

\begin{figure}
\includegraphics[width=\textwidth]{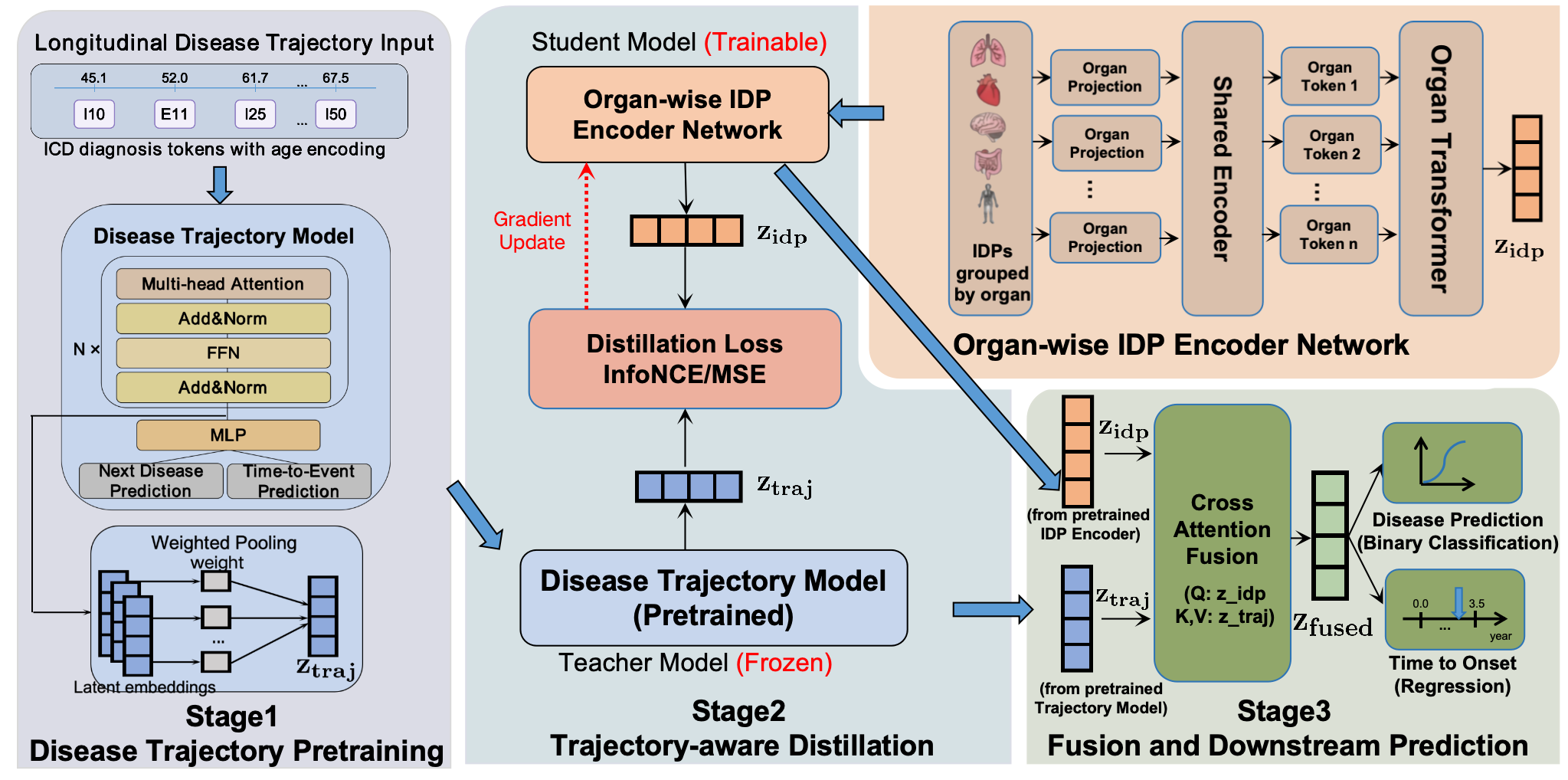}
\caption{
Trajectory-aware distillation framework for aligning imaging-derived phenotypes with disease trajectory representations.
(1) A generative disease trajectory Transformer learns population-scale disease progression structure and produces trajectory embeddings.
(2) An organ-wise IDP encoder is pretrained via representation alignment, encouraging imaging embeddings to preserve trajectory-informed geometry.
(3) Trajectory and imaging embeddings are optionally fused via cross-attention for downstream disease risk and time-to-onset prediction.
} \label{fig1}
\end{figure}
\section{Methods}
\subsection{Overview}
We aim to predict disease outcomes from multi-organ IDPs in realistic settings where imaging cohorts are limited, especially for low-prevalence diseases. We propose a \textbf{trajectory-aware distillation framework} that transfers population-level disease progression knowledge from large-scale longitudinal diagnosis data into an IDP-based discriminative model. The framework has three stages:
(1) \textbf{Trajectory pretraining} with a generative transformer on longitudinal disease sequences;
(2) \textbf{Trajectory-aware distillation} to align IDP and trajectory representations;
(3) \textbf{Fusion \& fine-tuning} for downstream prediction, optionally using cross-attention to integrate trajectory information.

\subsection{Generative Disease Trajectory Model}
We adopt a Delphi-style~\cite{shmatko2025} generative trajectory Transformer implemented as a GPT-2-like \emph{decoder-only} model.
Given an ordered sequence of diagnosis tokens $(c_1,\dots,c_T)$ with corresponding event ages $(a_1,\dots,a_T)$ (in days), the model uses causal self-attention with padding masking and predicts the next diagnosis event.

\textbf{Age encoding.}
To model continuous time, we add a sinusoidal age encoding to token embeddings: $x_t = \mathrm{Emb}(c_t) + \mathrm{AgeEnc}(a_t)$, where $\mathrm{AgeEnc}(\cdot)$ applies sine/cosine features followed by a linear projection.

\textbf{Training objective.}
The model is trained with a joint objective consisting of next-event cross entropy and a time-to-next-event likelihood term that models the distribution of inter-event time gaps.

\textbf{Trajectory embedding.}
After pretraining, the trajectory model is frozen and used as a teacher. For each subject, we compute a fixed-dimensional trajectory representation $z_i^{traj}\in\mathbb{R}^{d_t}$ by pooling the final-layer hidden states over non-padding positions (e.g., weighted/mean pooling), which serves as the supervision signal for distillation.
\subsection{Organ-wise IDP Encoder}
Each subject has multi-organ IDPs $x_i=\{x_i^{(o)}\}_{o=1}^{O}$ with heterogeneous dimensions $x_i^{(o)}\in\mathbb{R}^{d_o}$. We project each organ into a shared space and inject organ identity:
\begin{equation}
h_i^{(o)}=W_o x_i^{(o)}+b_o,\quad h_i^{(o)}\in\mathbb{R}^{d}
\end{equation}
\begin{equation}
z_i^{(o)}=\mathrm{MLP}_{shared}\big([h_i^{(o)} \,\|\, e_o]\big),
\end{equation}
where $e_o\in\mathbb{R}^{d_e}$ is a learnable organ embedding. We then aggregate organ tokens with an \textbf{Organ Transformer}:
\begin{equation}
Z_i=[z_i^{cls}, z_i^{(1)},\dots,z_i^{(O)}]\in\mathbb{R}^{(O+1)\times d},
\end{equation}
\begin{equation}
\mathrm{Attn}(Q,K,V)=\mathrm{softmax}\left(\frac{QK^\top}{\sqrt{d}}\right)V,
\end{equation}
where $Q,K,V$ are linear projections of $Z_i$. The output $z_i^{cls}$ summarizes multi-organ phenotypes and is used for distillation and downstream prediction.
\subsection{Trajectory-aware Distillation Pretraining}

For subjects with both IDPs and longitudinal records, we align IDP and trajectory embeddings via lightweight projection heads:
\begin{equation}
\tilde{z}_i^{idp}=f_{idp}(z_i^{idp}),\quad 
\tilde{z}_i^{traj}=f_{traj}(z_i^{traj}),
\end{equation}
where $f_{idp}(\cdot)$ and $f_{traj}(\cdot)$ are MLPs with matched output dimensionality.

\paragraph{Contrastive alignment (InfoNCE).}
Our default objective is the InfoNCE loss \cite{oord2018cpc,chen2020simclr}, which encourages matched IDP-trajectory pairs to be more similar than mismatched pairs:
\begin{equation}
\mathcal{L}_{distill}^{InfoNCE}
=
-\frac{1}{N}
\sum_{i=1}^{N}
\log
\frac{
\exp(\mathrm{sim}(\tilde{z}_i^{idp},\tilde{z}_i^{traj})/\tau)
}{
\sum_{j=1}^{N}
\exp(\mathrm{sim}(\tilde{z}_i^{idp},\tilde{z}_j^{traj})/\tau)
},
\end{equation}
where $\mathrm{sim}(\cdot,\cdot)$ denotes cosine similarity and $\tau$ is a temperature parameter, default set to 0.07.

\paragraph{Comparison with MSE alignment.}
We also evaluate a mean squared error (MSE) objective:
\begin{equation}
\mathcal{L}_{distill}^{MSE}
=
\frac{1}{N}
\sum_{i=1}^{N}
\left\|
\tilde{z}_i^{idp}-\tilde{z}_i^{traj}
\right\|_2^2.
\end{equation}

We evaluate both MSE and InfoNCE alignment objectives. Unless otherwise specified, InfoNCE is used in the main experiments, as it provides stronger ranking performance (AUC), while MSE tends to favor regression metrics (MAE).

\subsection{Cross-attention Fusion and Downstream Prediction}
For downstream disease prediction, we optionally incorporate trajectory information via cross-attention. We use organ-level IDP tokens as queries and trajectory embeddings as key--value tokens, enabling organ-specific phenotypes to attend to trajectory patterns relevant for prediction. The fused representation is then used for disease risk and time-to-onset prediction during fine-tuning.

\section{Experiments}

\begin{table}[t]
\centering
\caption{
Effect of trajectory-aware pretraining on different architectures. For each model, we report scratch training and the best-performing distillation objective (MSE or InfoNCE, selected by validation AUC). Results are averaged across 159 diseases. Improvements are significant under paired Wilcoxon signed-rank tests across diseases ($p<1e{-4}$).
}
\label{tab:model_performance}
\begin{tabular}{|l|l|c|c|}
\hline
\textbf{Model} & \textbf{Setting} & \textbf{AUC $\uparrow$} & \textbf{MAE $\downarrow$} \\
\hline
Fusion (XAttn) & Scratch & 0.6521 & 1.6407 \\
               & Pretrained (InfoNCE) & \textbf{0.6637} & \textbf{1.6019} \\
\hline
Fusion (Concat) & Scratch & 0.6246 & 1.6597 \\
                & Pretrained (InfoNCE) & \textbf{0.6611} & \textbf{1.5566} \\
\hline
IDP Organ (Transformer) & Scratch & 0.6065 & 1.7038 \\
                        & Pretrained (MSE) & \textbf{0.6358} & \textbf{1.5916} \\
\hline
IDP Baseline & Scratch & 0.6131 & 1.7373 \\
             & Pretrained (InfoNCE) & \textbf{0.6462} & \textbf{1.5997} \\
\hline
Trajectory (Weighted Pool) & -- & 0.6322 & \textbf{1.5523} \\
\hline
\end{tabular}
\end{table}

\noindent\textbf{Dataset and Experimental Setup.}
We use the UK Biobank (UKB) cohort with 502,387 participants and linked longitudinal diagnosis and imaging data. We select 301 quantitative phenotypes grouped into five system-level input sets: brain MRI regional grey-matter volumes; skeleton/body-composition measures; cardiac MRI traits; abdominal MRI traits; and pulmonary measures including spirometry indices and lung volume. After quality control, 35,193 participants with complete IDPs are retained. IDPs with missing rate $>20\%$ are removed and remaining missing values are median-imputed.

The cohort is partitioned into three \textbf{mutually exclusive} stages to prevent information leakage:

(1) \textbf{Trajectory pretraining}: 467,194 participants with diagnosis records only and no imaging data. The generative trajectory teacher is trained exclusively on this subset.

(2) \textbf{Distillation}: 15,000 participants with both IDPs and longitudinal trajectories, \textbf{used solely for representation alignment} between IDP and trajectory embeddings.

(3) \textbf{Downstream prediction}: 20,193 IDP participants for downstream prediction across 159 diseases ($\geq 40$ cases). These subjects are \textbf{never used} during trajectory pretraining or distillation.

For downstream evaluation, we use an \textbf{incident-only risk-set design}: imaging serves as the baseline time point, subjects with target-disease onset at or before imaging are excluded, and controls for each case are sampled from subjects who were already imaged and still event-free at the case time. Thus, the task is future-onset prediction after imaging rather than current-state classification.

To further prevent temporal leakage, trajectory embeddings for both distillation and evaluation subjects are extracted using diagnosis records \textbf{strictly prior to} imaging acquisition. Consequently, the fusion model uses historical pre-imaging EHR context, but does not access future diagnoses or post-imaging target information.

Unless specified, we use the Organ Transformer for IDP encoding, weighted pooling for trajectory embeddings, and cross-attention for fusion. The \textbf{IDP Baseline} predicts from concatenated IDPs, \textbf{IDP Organ (Transformer)} is the proposed organ-token encoder, \textbf{Fusion (Concat/XAttn)} combines IDP and trajectory representations, and \textbf{Trajectory (Weighted Pool)} uses only the pretrained EHR trajectory embedding. All results are averaged over 10 seeds.

\noindent\textbf{Effect of Trajectory-aware Pretraining.}
Table~\ref{tab:model_performance} reports scratch results and the best-performing alignment objective for each architecture. Trajectory-aware pretraining consistently improves IDP-based prediction across all architectures. For example, the Organ Transformer improves mean AUC from 0.6065 to 0.6358 (+0.0293), and the IDP baseline improves from 0.6131 to 0.6462 (+0.0331).

\noindent\textbf{Comparison of Alignment Objectives.}
We evaluate both MSE and InfoNCE objectives for trajectory-aware alignment. Both outperform scratch training but induce different tradeoffs: MSE tends to favor MAE, whereas InfoNCE better favors ranking-based discrimination (AUC). This suggests that the main benefit comes from trajectory-informed structural alignment rather than one specific loss.

\noindent\textbf{Fusion with Trajectory Embeddings.}
Integrating trajectory embeddings at fine-tuning further improves performance. Cross-attention fusion achieves the best overall AUC (0.6637), indicating complementary information between longitudinal and imaging representations.

\begin{figure*}[t]
\centering
\includegraphics[width=\textwidth]{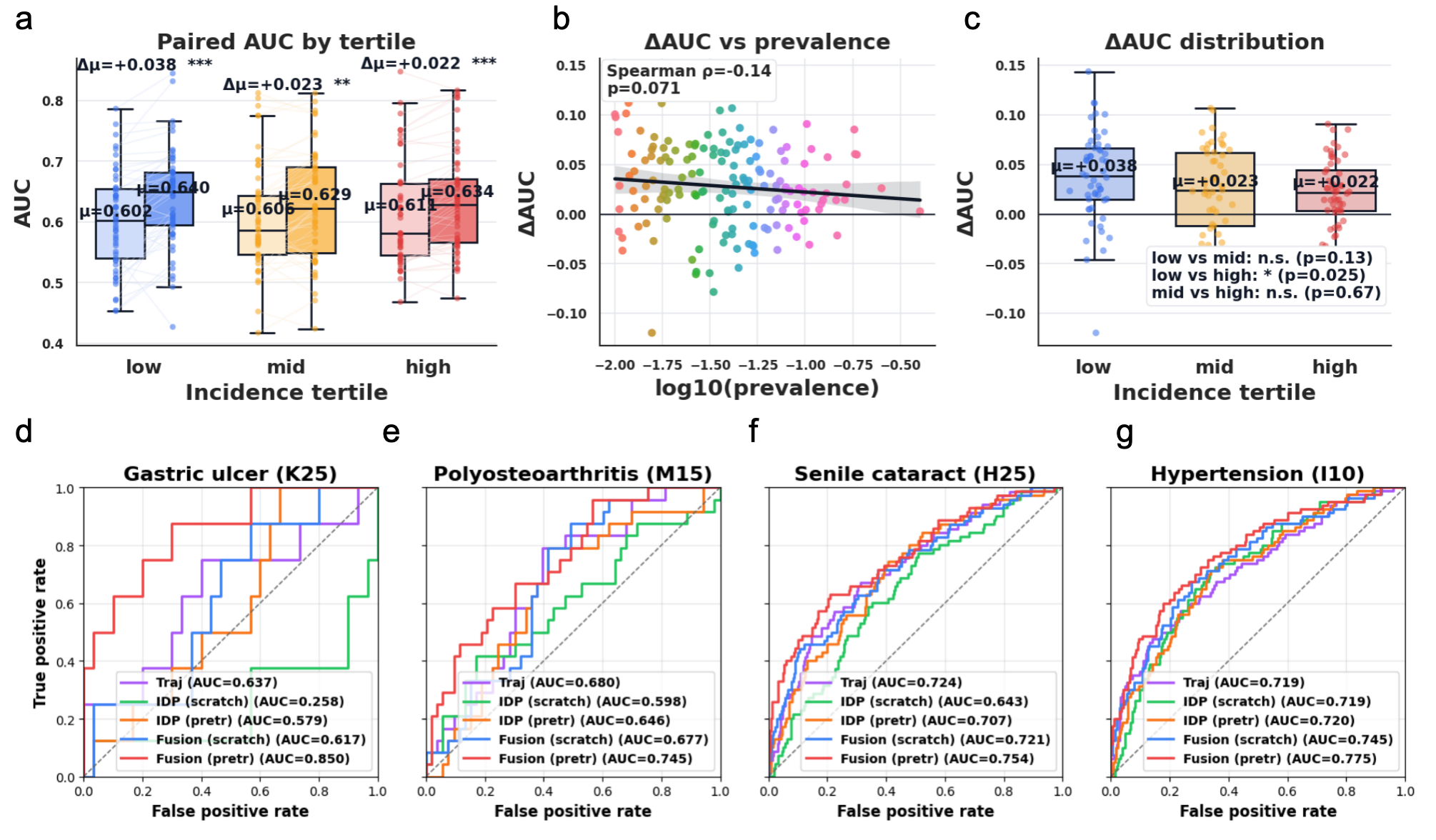}
\caption{
Effect of trajectory-aware pretraining across disease prevalence levels. 
Diseases are stratified into low-, mid-, and high-prevalence tertiles by evenly partitioning the 159 diseases based on prevalence.
(a) Paired AUC comparison across prevalence tertiles. 
(b) Relationship between prevalence and performance gain. 
(c) Distribution of $\Delta$AUC across tertiles. 
(d-g) ROC curves for representative diseases covering different prevalence levels. 
IDP uses the Organ Transformer, trajectory uses weighted pooling, and fusion uses cross-attention.
}
\label{fig:prevalence_analysis}
\end{figure*}

\noindent\textbf{Effect Across Disease Prevalence.}
We stratify diseases into tertiles based on prevalence to evaluate robustness across disease frequencies (Fig.~\ref{fig:prevalence_analysis}). 

Trajectory-aware pretraining significantly improves performance across all prevalence levels. The largest improvement is observed in the low-prevalence tertile (+0.0379 AUC), compared with +0.023 and +0.022 in the mid- and high-prevalence tertiles; the gain for rare diseases is significantly greater than that for common diseases (Mann--Whitney test, $p=0.0249$). Spearman correlation further reveals a negative association between disease prevalence and performance gain ($\rho=-0.14$, $p=0.071$), suggesting a trend toward larger benefits under lower-prevalence settings.

Representative ROC curves (Fig.~\ref{fig:prevalence_analysis}d-g) further demonstrate that trajectory-aware pretraining improves IDP models across diverse diseases, while fusion achieves the strongest overall performance, confirming the complementary roles of trajectory and imaging representations.

\begin{figure}[t]
\centering
\includegraphics[width=\textwidth]{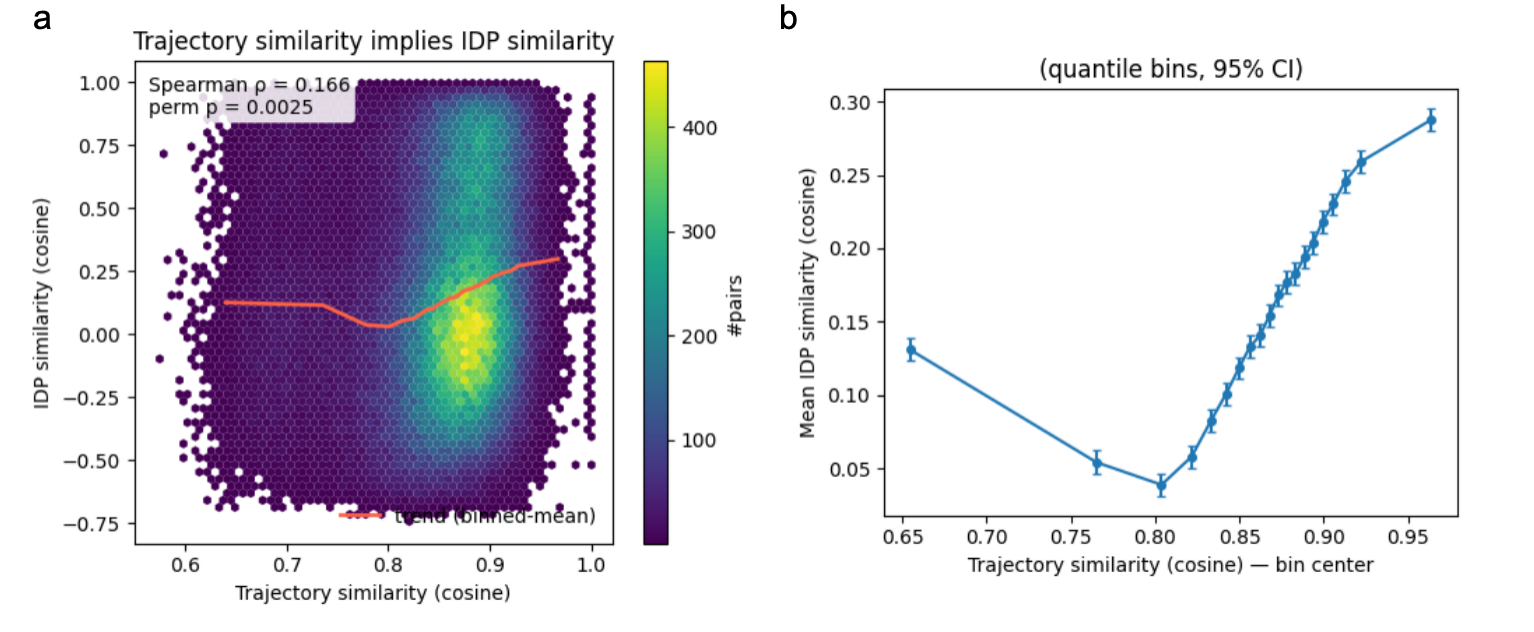}
\caption{ Representation geometry alignment between IDPs and disease trajectories. (a) Pairwise cosine similarity between trajectory embeddings and IDP embeddings across sampled subject pairs. Each hexagon indicates pair density, and the red curve shows the binned mean trend. Higher trajectory similarity corresponds to higher IDP similarity, indicating preservation of structural relationships. (b) Mean IDP similarity across quantile bins of trajectory similarity (error bars: 95\% confidence intervals), demonstrating monotonic geometric alignment between imaging and trajectory representation spaces. }
\label{fig2}
\end{figure}

\noindent\textbf{Representation Geometry Alignment.}
We examine whether similarity in imaging-derived phenotypes reflects similarity in disease trajectories. Fig.~\ref{fig2} shows a positive association between pairwise similarity in IDP and trajectory embeddings, with significant Spearman correlation ($\rho=0.166$, permutation $p=0.0025$). The binned analysis in Fig.~\ref{fig2}(b) shows the same monotonic trend despite substantial within-bin variability. These results support the effectiveness of trajectory-aware distillation.

\section{Conclusion}

We propose a trajectory-aware distillation framework that transfers structural disease progression knowledge from generative trajectory models into imaging-derived phenotype representations. By aligning IDP embeddings with trajectory embeddings, our approach improves robustness and predictive performance under limited imaging supervision.

Beyond performance improvements, our results provide empirical evidence that imaging-derived phenotypes and longitudinal disease trajectories share a common latent representation geometry. This suggests that generative models trained on large-scale longitudinal health records can serve as scalable structural priors for static biomedical modalities.

This work establishes a new paradigm for medical representation learning, where population-scale generative models guide representation learning in data-limited imaging settings.

Several limitations should be noted. First, all experiments are conducted on a single cohort, so the generalizability of the proposed framework remains to be established on external datasets. Second, our experimental comparisons are primarily designed to isolate the effect of trajectory-informed structural priors through controlled IDP-only, trajectory-only, and fusion comparisons, rather than to provide an exhaustive benchmark against the full landscape of EHR and multimodal disease-prediction methods. Stronger comparisons to alternative trajectory backbones and broader multimodal baselines would further strengthen the empirical positioning of the method. Third, we use one generative disease trajectory model as the teacher; it remains unclear how alternative trajectory models such as BEHRT or Med-BERT would behave in the same teacher role. Future work will therefore evaluate external cohorts, compare different disease trajectory models as teachers, broaden baseline coverage, and investigate more effective ways to align disease trajectory information with imaging phenotypes. Another promising direction is to use trajectory-informed structure not only for supervision, but also as an evaluation signal for measuring the disease relevance of imaging phenotypes or related representations.

\begin{credits}
\subsubsection{\ackname} This research was conducted using the UK Biobank Resource under Application Number 96511. We wish to thank all UK Biobank participants and staff. This work was supported in part by the Shanghai Rising-Star Program (No.24QA2703300), the Shanghai Municipal Science and Technology Major Project (No.2023SHZDZX02A05), the National Natural Science Foundation of China (No.62331021), the Scientific Research Fund Project of Pudong Hospital Affiliated to Fudan University (No.YJJC202409), and the National Key R\&D Program of China (No.2024YFC3405800). We acknowledge the Human Phenome Institute of Fudan University and the Multimodal Imaging Platform for providing essential resources and technical assistance. The computations in this research were performed using the CFFF platform of Fudan University.

\subsubsection{\discintname}
The authors have no competing interests to declare that are relevant to the content of this article.
\end{credits}

%
%
%

\begin{thebibliography}{8}

\bibitem{bycroft2018}
Bycroft, C., Freeman, C., Petkova, D. et al.: The UK Biobank resource with deep phenotyping and genomic data. Nature 562, 203--209 (2018). \doi{10.1038/s41586-018-0579-z}

\bibitem{shmatko2025}
Shmatko, A., Jung, A.W., Gaurav, K., Brunak, S., Mortensen, L.H., Birney, E., Fitzgerald, T., Gerstung, M.: Learning the natural history of human disease with generative transformers. Nature 647, 248-256 (2025). \doi{10.1038/s41586-025-09529-3}

\bibitem{miller2016}
Miller, K.L., Alfaro-Almagro, F., Bangerter, N.K. et al.: Multimodal population brain imaging in the UK Biobank prospective epidemiological study. Nature Neuroscience 19, 1523--1536 (2016). \doi{10.1038/nn.4393}

\bibitem{li2020behrt}
Li, Y., Rao, S., Solares, J.R.A. et al.: BEHRT: Transformer for electronic health records. Scientific Reports 10, 7155 (2020). \doi{10.1038/s41598-020-62922-y}

\bibitem{rasmy2021medbert}
Rasmy, L., Xiang, Y., Xie, Z. et al.: Med-BERT: pretrained contextualized embeddings on large-scale structured electronic health records for disease prediction. npj Digital Medicine 4, 86 (2021). \doi{10.1038/s41746-021-00455-y}

\bibitem{bengio2013representation}
Bengio, Y., Courville, A., Vincent, P.: Representation learning: A review and new perspectives. IEEE Transactions on Pattern Analysis and Machine Intelligence 35(8), 1798--1828 (2013). \doi{10.1109/TPAMI.2013.50}

\bibitem{radford2021clip}
Radford, A., Kim, J.W., Hallacy, C. et al.: Learning transferable visual models from natural language supervision. In: Proceedings of the International Conference on Machine Learning (ICML), pp. 8748--8763 (2021)

\bibitem{liu2025multimodal}
Liu, C., Ye, F.: A review of multimodal medical data fusion techniques for personalized medicine. In: Proceedings of the 4th International Conference on Biomedical and Intelligent Systems (IC-BIS), pp. 338--347 (2025). \doi{10.1145/3745034.3745088}

\bibitem{hinton2015distillation}
Hinton, G., Vinyals, O., Dean, J.: Distilling the knowledge in a neural network. arXiv preprint arXiv:1503.02531 (2015)

\bibitem{oord2018cpc}
Oord, A.v.d., Li, Y., Vinyals, O.: Representation learning with contrastive predictive coding. arXiv preprint arXiv:1807.03748 (2018)

\bibitem{chen2020simclr}
Chen, T., Kornblith, S., Norouzi, M., Hinton, G.: A simple framework for contrastive learning of visual representations. In: Proceedings of the International Conference on Machine Learning (ICML), pp. 1597--1607 (2020)

\bibitem{tian2020crd}
Tian, Y., Krishnan, D., Isola, P.: Contrastive representation distillation. In: Proceedings of the International Conference on Learning Representations (ICLR) (2020)

\end{thebibliography}
%

\end{document}